\pdfoutput=1
\documentclass[12pt,a4paper]{article}
\usepackage{ifthen} %
\newboolean{pdflatex}
\setboolean{pdflatex}{true} %
\usepackage{overpic} 

\newboolean{articletitles}
\setboolean{articletitles}{true} %

\newboolean{uprightparticles}
\setboolean{uprightparticles}{false} %

\usepackage{color}

\newcommand{\Dpp}{\ensuremath{\Dz\to \pip\pim}\xspace}
\newcommand{\Dkk}{\ensuremath{\Dz\to\Kp\Km}\xspace}
\newcommand{\Dkp}{\ensuremath{\Dz\to\Km\pip}\xspace}

\newcommand{\mD}{\ensuremath{m(\Dz)}\xspace}

\newcommand{\Units}{10^{-4}}
\newcommand{\ppRes}{2.2}
\newcommand{\ppStat}{7.0}
\newcommand{\ppSyst}{0.8}

\newcommand{\kkRes}{-4.3}
\newcommand{\kkStat}{3.6}
\newcommand{\kkSyst}{0.5}

\newcommand{\kpRes}{1.6}
\newcommand{\kpStat}{1.2}

\newcommand{\kkCombRes}{-4.4}
\newcommand{\kkCombStat}{2.3}
\newcommand{\kkCombSyst}{0.6}

\newcommand{\ppCombRes}{2.5}
\newcommand{\ppCombStat}{4.3}
\newcommand{\ppCombSyst}{0.7}

\def\paperauthors{LHCb collaboration} %
\def\paperasciititle{Updated measurement of decay-time-dependent CP asymmetries in D0->K+K- and D0->pi+pi- decays} %
\def\papertitle{Updated measurement of decay-time-dependent \CP asymmetries in \Dkk and \Dpp decays} %
\def\paperkeywords{{High Energy Physics}, {LHCb}} %
\def\papercopyright{\the\year\ CERN for the benefit of the LHCb collaboration} %
\def\paperlicence{CC-BY-4.0 licence}
\def\paperlicenceurl{https://creativecommons.org/licenses/by/4.0/}

\def\mycernpapernumber{CERN-EP-2019-225}
\def\mylhcbpapernumber{LHCb-PAPER-2019-032}
\def\mydate{January 13, 2020}
\def\myversion{}
\usepackage[top=1in, bottom=1.25in, left=1in, right=1in]{geometry}

\usepackage{microtype}
\usepackage{lineno}  %
\usepackage{xspace} %
\usepackage{caption} %

\usepackage{graphicx}  %
\usepackage{color}
\usepackage{colortbl}
\graphicspath{{./figs/}} %
\DeclareGraphicsExtensions{.pdf,.PDF,png,.PNG}

\usepackage{amsmath} %
\usepackage{amssymb}
\usepackage{amsfonts}
\usepackage{upgreek} %

\newcommand*\patchAmsMathEnvironmentForLineno[1]{%
\expandafter\let\csname old#1\expandafter\endcsname\csname #1\endcsname
\expandafter\let\csname oldend#1\expandafter\endcsname\csname
end#1\endcsname
 \renewenvironment{#1}%
   {\linenomath\csname old#1\endcsname}%
   {\csname oldend#1\endcsname\endlinenomath}%
}
\newcommand*\patchBothAmsMathEnvironmentsForLineno[1]{%
  \patchAmsMathEnvironmentForLineno{#1}%
  \patchAmsMathEnvironmentForLineno{#1*}%
}
\AtBeginDocument{%
\patchBothAmsMathEnvironmentsForLineno{equation}%
\patchBothAmsMathEnvironmentsForLineno{align}%
\patchBothAmsMathEnvironmentsForLineno{flalign}%
\patchBothAmsMathEnvironmentsForLineno{alignat}%
\patchBothAmsMathEnvironmentsForLineno{gather}%
\patchBothAmsMathEnvironmentsForLineno{multline}%
\patchBothAmsMathEnvironmentsForLineno{eqnarray}%
}

\usepackage{hyperxmp}

\usepackage[pdftex,
            pdfauthor={\paperauthors},
            pdftitle={\paperasciititle},
            pdfkeywords={\paperkeywords},
            pdfcopyright={Copyright (C) \papercopyright},
            pdflicenseurl={\paperlicenceurl}]{hyperref}

\usepackage[colorinlistoftodos,textsize=scriptsize]{todonotes}

\usepackage[all]{hypcap} %

\usepackage{xspace} 
\usepackage{upgreek}

\def\af {{\ensuremath{A_f}}\xspace}
\def\abf {{\ensuremath{\bar{A}_f}}\xspace}

\def\CPV {{\ensuremath{C\!PV}}\xspace}

\def\lhcb   {\mbox{LHCb}\xspace}

\def\babar  {\mbox{BaBar}\xspace}
\def\belle  {\mbox{Belle}\xspace}

\def\cdf    {\mbox{CDF}\xspace}

\def\MagUp {\mbox{\em Mag\kern -0.05em Up}\xspace}

\ifthenelse{\boolean{uprightparticles}}%
{

 \def\Pmu         {\ensuremath{\upmu}\xspace}

 \def\Ppi         {\ensuremath{\uppi}\xspace}

 \def\PDelta      {\ensuremath{\Delta}\xspace}                 
 \def\PXi         {\ensuremath{\Xi}\xspace}                 
 \def\PLambda     {\ensuremath{\Lambda}\xspace}                 
 \def\PSigma      {\ensuremath{\Sigma}\xspace}                 
 \def\POmega      {\ensuremath{\Omega}\xspace}                 
 \def\PUpsilon    {\ensuremath{\Upsilon}\xspace}

 \def\PB      {\ensuremath{\mathrm{B}}\xspace}                 
                  
 \def\PD      {\ensuremath{\mathrm{D}}\xspace}

 \def\PK      {\ensuremath{\mathrm{K}}\xspace}

 \def\Pb      {\ensuremath{\mathrm{b}}\xspace}                 
 \def\Pc      {\ensuremath{\mathrm{c}}\xspace}

 \def\Pi      {\ensuremath{\mathrm{i}}\xspace}

 \def\Ps      {\ensuremath{\mathrm{s}}\xspace}

 \def\thebaroffset{0.0em}
}
{

 \def\Pmu         {\ensuremath{\mu}\xspace}

 \def\Ppi         {\ensuremath{\pi}\xspace}

 \mathchardef\PDelta="7101
 \mathchardef\PXi="7104
 \mathchardef\PLambda="7103
 \mathchardef\PSigma="7106
 \mathchardef\POmega="710A
 \mathchardef\PUpsilon="7107
                  
 \def\PB      {\ensuremath{B}\xspace}                 
                  
 \def\PD      {\ensuremath{D}\xspace}

 \def\PK      {\ensuremath{K}\xspace}

 \def\Pb      {\ensuremath{b}\xspace}                 
 \def\Pc      {\ensuremath{c}\xspace}

 \def\Pi      {\ensuremath{i}\xspace}

 \def\Ps      {\ensuremath{s}\xspace}

 \def\thebaroffset{0.18em}
}
\newcommand{\offsetoverline}[2][\thebaroffset]{\kern #1\overline{\kern -#1 #2}}%

\makeatletter
\ifcase \@ptsize \relax%
  \newcommand{\miniscule}{\@setfontsize\miniscule{4}{5}}%
\or%
  \newcommand{\miniscule}{\@setfontsize\miniscule{5}{6}}%
\or%
  \newcommand{\miniscule}{\@setfontsize\miniscule{5}{6}}%
\fi
\makeatother

\DeclareRobustCommand{\optbar}[1]{\shortstack{{\miniscule (\rule[.5ex]{1.25em}{.18mm})}
  \\ [-.7ex] $#1$}}

\def\mun        {{\ensuremath{\Pmu^-}}\xspace} %

\def\squark    {{\ensuremath{\Ps}}\xspace}

\def\cquark    {{\ensuremath{\Pc}}\xspace}

\def\bquark    {{\ensuremath{\Pb}}\xspace}

\def\pion   {{\ensuremath{\Ppi}}\xspace}

\def\pip    {{\ensuremath{\pion^+}}\xspace}
\def\pim    {{\ensuremath{\pion^-}}\xspace}

\def\kaon    {{\ensuremath{\PK}}\xspace}

\def\KorKbar {\kern \thebaroffset\optbar{\kern -\thebaroffset \PK}{}\xspace}

\def\Kp      {{\ensuremath{\kaon^+}}\xspace}
\def\Km      {{\ensuremath{\kaon^-}}\xspace}
\def\Kpm     {{\ensuremath{\kaon^\pm}}\xspace}

\def\Dbar    {{\ensuremath{\offsetoverline{\PD}}}\xspace}
\def\D       {{\ensuremath{\PD}}\xspace}

\def\DorDbar {\kern \thebaroffset\optbar{\kern -\thebaroffset \PD}\xspace}
\def\Dz      {{\ensuremath{\D^0}}\xspace}
\def\Dzb     {{\ensuremath{\Dbar{}^0}}\xspace}

\def\Dstarp  {{\ensuremath{\D^{*+}}}\xspace}

\def\B       {{\ensuremath{\PB}}\xspace}
\def\Bbar    {{\ensuremath{\offsetoverline{\PB}}}\xspace}

\def\BorBbar {\kern \thebaroffset\optbar{\kern -\thebaroffset \PB}\xspace}

\def\Bzb     {{\ensuremath{\Bbar{}^0}}\xspace}
\def\Bd      {{\ensuremath{\B^0}}\xspace}

\def\BdorBdbar {\kern \thebaroffset\optbar{\kern -\thebaroffset \Bd}\xspace}

\def\Bub     {{\ensuremath{\B^-}}\xspace}

\def\Bm      {{\ensuremath{\Bub}}\xspace}
\def\Bpm     {{\ensuremath{\B^\pm}}\xspace}

\def\Bs      {{\ensuremath{\B^0_\squark}}\xspace}

\def\BsorBsbar {\kern \thebaroffset\optbar{\kern -\thebaroffset \Bs}\xspace}

\def\Y#1S{\ensuremath{\PUpsilon{(#1S)}}\xspace}

\def\LorLbar     {\kern \thebaroffset\optbar{\kern -\thebaroffset \PLambda}\xspace}

\newcommand{\decay}[2]{\ensuremath{#1\!\to #2}\xspace} 

\def\to                 {\ensuremath{\rightarrow}\xspace}

\def\CP                {{\ensuremath{C\!P}}\xspace}

\newcommand{\ACP}{{\ensuremath{{\mathcal{A}}_{\CP}}}\xspace}
\newcommand{\ACPdir}{{\ensuremath{{\mathcal{A}}^\text{dir}_{\CP}}}\xspace}

\newcommand{\Araw}{{\ensuremath{A_\text{raw}}}\xspace}

\def\AT#1     {\ensuremath{A_{\mathrm{T}}^{#1}}\xspace}           %

\def\C#1      {\ensuremath{\mathcal{C}_{#1}}\xspace}                       %
\def\Cp#1     {\ensuremath{\mathcal{C}_{#1}^{'}}\xspace}                    %
\def\Ceff#1   {\ensuremath{\mathcal{C}_{#1}^{\mathrm{(eff)}}}\xspace}        %
\def\Cpeff#1  {\ensuremath{\mathcal{C}_{#1}^{'\mathrm{(eff)}}}\xspace}       %
\def\Ope#1    {\ensuremath{\mathcal{O}_{#1}}\xspace}                       %
\def\Opep#1   {\ensuremath{\mathcal{O}_{#1}^{'}}\xspace}                    %

\def\agamma     {\ensuremath{A_{\Gamma}}\xspace}
\newcommand{\ket}[1]{\ensuremath{|#1\rangle}}              %
\newcommand{\nospaceunit}[1]{\ensuremath{\text{#1}}}       
\newcommand{\aunit}[1]{\ensuremath{\text{\,#1}}}       
\newcommand{\tev}{\aunit{Te\kern -0.1em V}\xspace}
\newcommand{\gev}{\aunit{Ge\kern -0.1em V}\xspace}
\newcommand{\mev}{\aunit{Me\kern -0.1em V}\xspace}
\newcommand{\kev}{\aunit{ke\kern -0.1em V}\xspace}
\newcommand{\ev}{\aunit{e\kern -0.1em V}\xspace}
\newcommand{\mevc}{\ensuremath{\aunit{Me\kern -0.1em V\!/}c}\xspace}
\newcommand{\gevc}{\ensuremath{\aunit{Ge\kern -0.1em V\!/}c}\xspace}
\newcommand{\mevcc}{\ensuremath{\aunit{Me\kern -0.1em V\!/}c^2}\xspace}
\newcommand{\gevcc}{\ensuremath{\aunit{Ge\kern -0.1em V\!/}c^2}\xspace}

\def\mum  {\ensuremath{\,\upmu\nospaceunit{m}}\xspace}

\def\fb   {\ensuremath{\aunit{fb}}\xspace}
\def\invfb   {\ensuremath{\fb^{-1}}\xspace}

\def\fs   {\aunit{fs}}

\def\gsim{{~\raise.15em\hbox{$>$}\kern-.85em
          \lower.35em\hbox{$\sim$}~}\xspace}
\def\lsim{{~\raise.15em\hbox{$<$}\kern-.85em
          \lower.35em\hbox{$\sim$}~}\xspace}

\def\pt         {\ensuremath{p_{\mathrm{T}}}\xspace}

\def\ptot       {\ensuremath{p}\xspace}

\def\tell1  {TELL1\xspace}
\def\ukl1   {UKL1\xspace}

\usepackage{cite} %
\usepackage{mciteplus}
\usepackage{booktabs}
 \usepackage{longtable} %

\begin{document}

\renewcommand{\thefootnote}{\fnsymbol{footnote}}
\setcounter{footnote}{1}

%
%
%
%
%
%
%
%
%

%
%
%
\begin{titlepage}
\pagenumbering{roman}

\vspace*{-1.5cm}
\centerline{\large EUROPEAN ORGANIZATION FOR NUCLEAR RESEARCH (CERN)}
\vspace*{1.5cm}
\noindent
\begin{tabular*}{\linewidth}{lc@{\extracolsep{\fill}}r@{\extracolsep{0pt}}}
\ifthenelse{\boolean{pdflatex}}%
{\vspace*{-1.5cm}\mbox{\!\!\!\includegraphics[width=.14\textwidth]{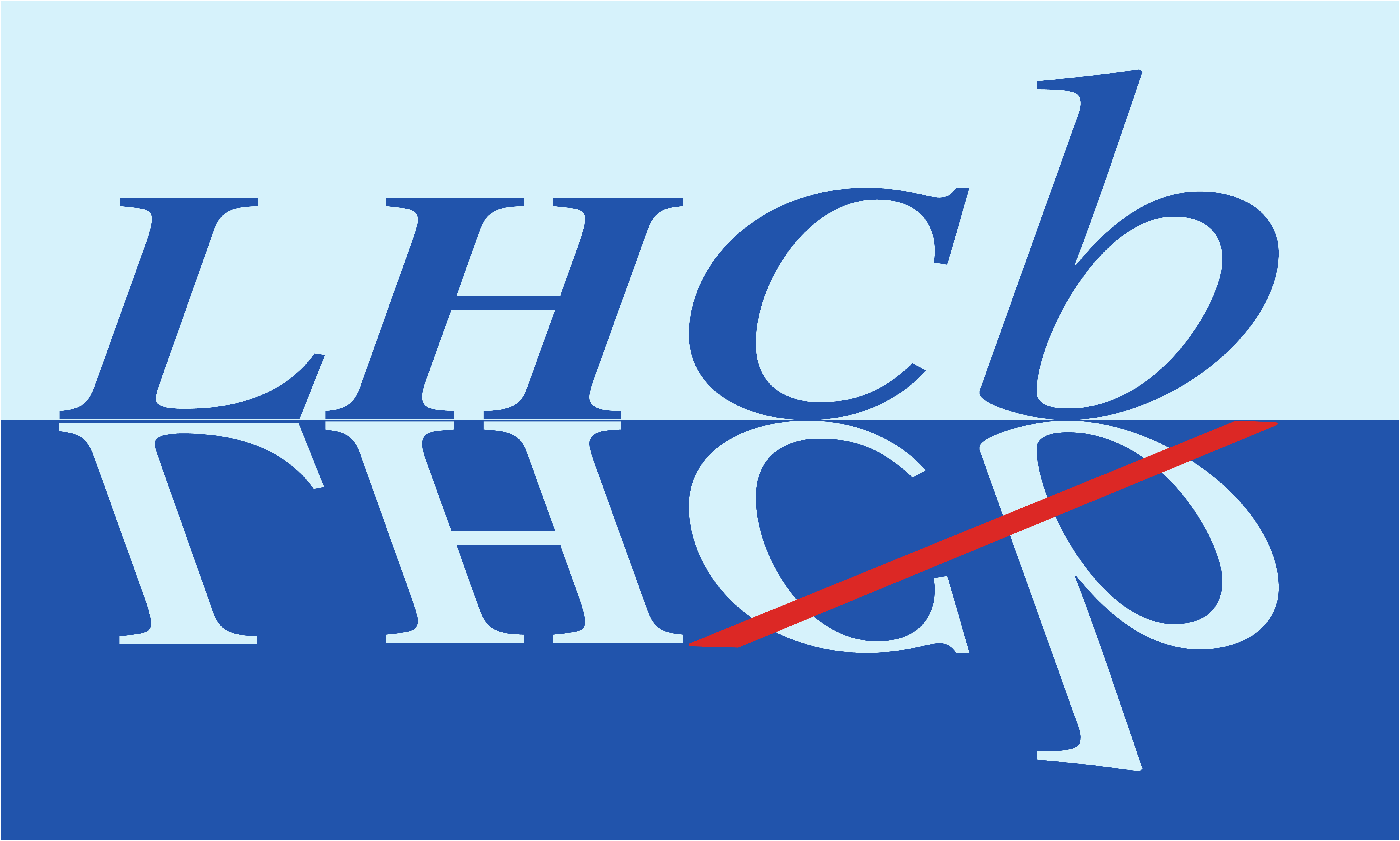}} & &}%
{\vspace*{-1.2cm}\mbox{\!\!\!\includegraphics[width=.12\textwidth]{lhcb-logo.eps}} & &}%
\\
 & & \mycernpapernumber \\  %
 & & \mylhcbpapernumber \\  %
 & & \mydate \\ %
 & & \myversion \\
\end{tabular*}

\vspace*{3.0cm}

{\normalfont\bfseries\boldmath\huge
\begin{center}
  \papertitle 
\end{center}
}

\vspace*{2.0cm}

\begin{center}
\paperauthors\footnote{Authors are listed at the end of this paper.}
\end{center}

\vspace{\fill}

\begin{abstract}
%
\noindent A search for decay-time-dependent charge-parity (\CP) asymmetry in \mbox{\Dkk} and \mbox{\Dpp} decays is performed at the \lhcb experiment using proton-proton collision data recorded at a center-of-mass energy of 13\tev, and corresponding to an integrated luminosity of 5.4\invfb. The \Dz mesons are required to originate from semileptonic decays of \bquark hadrons, such that the charge of the muon identifies the flavor of the neutral \D meson at production. The asymmetries in the effective decay widths of \Dz and \Dzb mesons are determined to be \mbox{$\agamma(\Kp\Km) = (\kkRes \pm \kkStat  \pm \kkSyst)\times\Units$} and \mbox{$\agamma(\pip\pim)=(\ppRes \pm \ppStat  \pm \ppSyst)\times\Units$}, where the uncertainties are statistical and systematic, respectively. The results are consistent with \CP symmetry and, when combined with previous \lhcb results, yield \mbox{$\agamma(\Kp\Km) = (\kkCombRes \pm \kkCombStat  \pm \kkCombSyst)\times\Units$} and \mbox{$\agamma(\pip\pim)=(\ppCombRes \pm \ppCombStat  \pm \ppCombSyst)\times\Units$}.
 \end{abstract}

\vspace*{1.0cm}

\begin{center}
        Published in Phys. Rev. D101 (2020) 012005
\end{center}

\vspace{\fill}

{\footnotesize 
\centerline{\copyright~\papercopyright. \href{\paperlicenceurl}{\paperlicence}.}}
\vspace*{2mm}

\end{titlepage}

\newpage
\setcounter{page}{2}
\mbox{~}

\cleardoublepage

 %
%

\renewcommand{\thefootnote}{\arabic{footnote}}
\setcounter{footnote}{0}

\pagestyle{plain} %
\setcounter{page}{1}
\pagenumbering{arabic}

%
%
%

%
\section{Introduction}
Charge-parity (\CP) violation is one of the key ingredients that are needed to generate the asymmetry between matter and antimatter observed in the Universe~\cite{Sakharov:1967dj}. The Standard Model (SM) of particle physics, where all known \CP-violating processes arise from the irreducible phase of the Cabibbo-Kobayashi-Maskawa matrix~\cite{Cabibbo:1963yz,Kobayashi:1973fv} is ,however, unable to explain the observed asymmetry~\cite{Huet:1994jb,Dine:2003ax}. New dynamics that lead to a significant enhancement of \CP-violating processes are required, making searches for \CP violation a powerful probe for physics beyond the SM. Although \CP violation has been experimentally observed in the down-type quark sector with measurements of $K$ and \B mesons~\cite{Christenson:1964fg,Aubert:2004qm,Chao:2004mn,LHCb-PAPER-2013-018,LHCb-PAPER-2012-001}, no indication of new dynamics has been reported yet. Only recently has \CP violation been observed in the decay of charmed mesons~\cite{LHCb-PAPER-2019-006}. The limited precision of the SM predictions, together with the limited amount of experimental information available~\cite{HFLAV18}, is, however, not yet sufficient to establish whether the observed signal could be explained by the SM~\cite{Chala:2019fdb,Li:2019hho,Grossman:2019xcj,Soni:2019xko,Buccella:2019kpn,Cheng:2019ggx}. Additional searches for \CP violation in the charm sector, and particularly for more suppressed and yet-to-be-observed signs of \CP-violating effects induced by \Dz--\Dzb mixing, have unique potential to probe for the existence of beyond-the-SM dynamics, which couple preferentially to up-type quarks~\cite{Golden:1989qx,Buccella:1994nf,Bianco:2003vb,Grossman:2006jg,Artuso:2008vf,Khodjamirian:2017zdu}.
 
This paper reports a search for \CP violation in \Dz--\Dzb mixing, or in the interference between mixing and decay, through the measurement of the asymmetry between the effective decay widths, $\hat{\Gamma}$, of mesons initially produced as \Dz and \Dzb and decaying into the \CP-even final states $f = \Kp\Km$, $\pip\pim$:
\begin{equation}
\agamma(f) \equiv \frac{\hat{\Gamma}({\Dz \to f})-\hat{\Gamma}({\Dzb \to f})}{\hat{\Gamma}({\Dz \to f})+\hat{\Gamma}({\Dzb \to f})}.
\end{equation}
Several measurements of the parameter $\agamma(f)$ have been performed by the \babar~\cite{Lees:2012qh}, \cdf~\cite{Aaltonen:2014efa}, \belle~\cite{Staric:2015sta}, and \lhcb~\cite{LHCb-PAPER-2014-069,LHCb-PAPER-2016-063,LHCb-PAPER-2019-001} Collaborations, leading to the current world-average value of \mbox{$(-3.2 \pm 2.6)\times\Units$}~\cite{HFLAV18}, when neglecting differences between the \Dkk and \Dpp decays.\footnote{Throughout the paper, the inclusion of the charge-conjugate decay mode is implied unless otherwise stated.} The achieved sensitivity is still 1 order of magnitude larger than the theoretical predictions of \mbox{$\agamma \approx 3\times 10^{-5}$}~\cite{Cerri:2018ypt}. This paper updates the \lhcb measurements of Refs.~\cite{LHCb-PAPER-2014-069,LHCb-PAPER-2016-063,LHCb-PAPER-2019-001} using the data sample of proton-proton collisions collected at a center-of-mass energy of 13\tev during 2016--2018, and corresponding to an integrated luminosity of 5.4\invfb. The analysis is performed using \Dz mesons originating from semileptonic decays of \bquark hadrons, where the \bquark-hadron candidates are only partially reconstructed. The charge of the muon identifies (``tags'') the flavor of the \Dz meson at its production. The samples are dominated by \mbox{$\Bm \to \Dz \mun X$} and \mbox{$\Bzb  \to \Dz \mun X$} decays, where $X$ denotes any set of final-state particles that are not reconstructed.

The paper is structured as follows: the analysis strategy is described in Sec.~\ref{sec:overview}. The LHCb detector is sketched in Sec.~\ref{sec:detector}; Sec.~\ref{sec:selection} details the criteria used to select the signal and control samples; Sec.~\ref{sec:fit} describes the fit method, and its validation using \Dkp decays; the determination of the systematic uncertainties is outlined in Sec.~\ref{sec:systematics}, before concluding with the presentation of the final results in Sec.~\ref{sec:results}.

\section{Analysis strategy\label{sec:overview}}
Due to the weak interactions, the mass eigenstates of neutral charm mesons, $D_1$ and $D_2$, are a superposition of the flavor states, \Dz and \Dzb: \mbox{$\ket{D_{1,2}} \equiv p \ket{\Dz} \pm q \ket{\Dzb}$}, where $q$ and $p$ are complex coefficients satisfying \mbox{$\lvert p\rvert^2 + \lvert q\rvert^2 = 1$}. Hence, an originally produced \Dz meson can oscillate as a function of time into a \Dzb meson, and vice versa, before decaying. In the limit of \CP symmetry, $q$ equals $p$ and the oscillations are characterized by only two dimensionless parameters, $x \equiv  (m_1 - m_2)c^2/\Gamma$ and $y \equiv (\Gamma_1 - \Gamma_2)/ 2\Gamma$, where $m_{1(2)}$ and $\Gamma_{1(2)}$ are the mass and decay width of the \CP-even (odd) eigenstate $D_{1(2)}$, respectively, and $\Gamma \equiv (\Gamma_1 + \Gamma_2)/2$ is the average decay width~\cite{PDG2018}. The values of $x$ and $y$ have been measured to be of the order of 1\% or smaller \cite{HFLAV18}. In the presence of \CP violation, the mixing rates for mesons produced as \Dz and \Dzb differ, further enriching the phenomenology. As an example, indicating with \af (\abf) the decay amplitude of a \Dz (\Dzb) meson into the final state $f$, three different manifestations of \CP violation can be measured: (i) \CP violation in the decay if $\ACPdir(f)\equiv(|\af|^2-|\abf|^2)/(|\af|^2+|\abf|^2)$ differs from zero, (ii) \CP violation in mixing if $|q/p|$ differs from unity,  and (iii) \CP violation in the interference between mixing and decay if $\phi_f \equiv \arg[(q\abf)/(p\af)]$ differs from zero. The latter two can be accessed by measuring the decay-time-dependent \CP asymmetry
\begin{equation}\label{eq:acp}
\ACP(\Dz\to f; t) = \frac{\Gamma(\Dz(t) \to f) - \Gamma(\Dzb(t) \to f)}{\Gamma(\Dz(t) \to f) + \Gamma(\Dzb(t) \to f)}.
\end{equation}
In the limit of small mixing parameters, Eq.~\eqref{eq:acp} can be approximated as a linear function of decay time~\cite{Aaltonen:2011se,Gersabeck:2011xj},
\begin{equation}
\ACP(\Dz\to f; t) \approx \ACPdir(f) - \agamma(f)\ \frac{t}{\tau}\label{eq:approx-acp},
\end{equation}
where $\tau=1/\Gamma$ is the average lifetime of neutral \D mesons. The coefficient $\agamma(f)$ is related to the mixing and \CP-violation parameters by~\cite{LHCb-CONF-2019-001}
\begin{equation}
\agamma(f)\approx -x\phi_f + y\left(|q/p| - 1\right) - y\ACPdir(f).
\end{equation}
Contrarily to the measurement reported in Ref.~\cite{LHCb-PAPER-2019-006}, which is sensitive to \mbox{$\ACPdir(\Kp\Km)-\ACPdir(\pip\pim)$}, $\agamma(f)$ is mostly sensitive to \CP violation in mixing or in the interference between mixing and decay, because the term \mbox{$y \ACPdir(f) \leqslant 10^{-5}$}~\cite{HFLAV18} can be neglected at the current level of experimental precision. Moreover, neglecting the $\mathcal{O}(10^{-3})$ difference between the weak phases of the decay amplitudes to the \CP-even final states $\Kp\Km$ and $\pip\pim$, $\phi_f\approx\phi\equiv\arg(q/p)$ becomes universal and \agamma independent of $f$~\cite{Grossman:2006jg}.

Experimentally, the partial rate asymmetry of Eq.~\eqref{eq:acp} cannot be measured directly because of charge-asymmetric detection efficiencies and asymmetric production rates of \Dz and \Dzb mesons from semileptonic \bquark-hadron decays in proton-proton collisions. Instead, the ``raw'' asymmetry between the \Dz and \Dzb mesons yields,
\begin{equation}
\Araw(\Dz\to f) = \frac{N(\Bbar\to\Dz(\to f)\mu^-X)-N(\B\to\Dzb(\to f)\mu^+X)}{N(\Bbar\to\Dz(\to f)\mu^-X)+N(\B\to\Dzb(\to f)\mu^+X)},
\end{equation}
is measured as a function of decay time. Neglecting higher-order terms in the involved asymmetries, which are at most $\mathcal{O}(1\%)$, the raw asymmetry can be approximated as
\begin{equation}
\Araw(\Dz\to f; t) \approx \ACP(\Dz\to f; t) + A_{D}(\mu) + A_{P}(\D),\label{eq:araw}
\end{equation}
where $A_D(\mu)$ and $A_P(\D)$ are the nuisance asymmetries due to the detection efficiency of the tagging muon and to the production rates of the neutral \D mesons, respectively. The parameter \agamma corresponds to the slope of the decay-time-dependent raw asymmetry only if $A_D$ and $A_P$ are independent of decay time. In this analysis, a possible time dependence of $A_D$ and $A_P$ is considered as a source of systematic uncertainty. The analysis procedure is validated on data using a control sample of Cabibbo-favored \Dkp decays, whose size exceeds that of the \Dkk and \Dpp signal modes by approximately 1 order of magnitude, and where measured asymmetries can be attributed solely to instrumental effects because no \CP violation is expected. To avoid potential experimenter's bias, the measured values of $\agamma(\Kp\Km)$ and $\agamma(\pip\pim)$ remained unknown during the development of the analysis and were examined only after the analysis procedure and the evaluation of the systematic uncertainties were finalized.

\section{Detector\label{sec:detector}}  
The \lhcb detector~\cite{LHCb-DP-2008-001,LHCb-DP-2014-002} is a single-arm forward spectrometer covering the \mbox{pseudorapidity} range $2<\eta <5$, designed for the study of particles containing \bquark or \cquark quarks. The detector includes a high-precision tracking system consisting of a silicon-strip vertex detector surrounding the $pp$ interaction region, a large-area silicon-strip detector located upstream of a dipole magnet with a bending power of about $4{\mathrm{\,Tm}}$, and three stations of silicon-strip detectors and straw drift tubes placed downstream of the magnet. The tracking system provides a measurement of the momentum, \ptot, of charged particles with relative uncertainty that varies from 0.5\% at low momentum to 1.0\% at 200\gevc. The minimum distance of a track to a primary vertex (PV), the impact parameter, is measured with a  resolution of $(15+29/\pt)\mum$, where \pt is the component of the momentum transverse to the beam, in\,\gevc. Different types of charged hadrons are distinguished using information from two ring-imaging Cherenkov detectors. Photons, electrons, and hadrons are identified by a calorimeter system consisting of scintillating-pad and preshower detectors, an electromagnetic and a hadronic calorimeter. Muons are identified by a system composed of alternating layers of iron and multiwire proportional chambers. The magnetic-field polarity is reversed periodically during data taking to mitigate the differences of reconstruction efficiencies of particles with opposite charges.

The on-line event selection is performed by a trigger, which consists of a hardware stage followed by a two-level software stage. In between the two software stages, an alignment and calibration of the detector is performed in near real time~\cite{LHCb-PROC-2015-011}. The same alignment and calibration information is propagated to the off-line reconstruction, ensuring consistent and high-quality particle identification information between the trigger and off-line software. The identical performance of the on-line and off-line reconstruction offers the opportunity to perform physics analyses directly using candidates reconstructed in the trigger \cite{LHCb-DP-2012-004,LHCb-DP-2016-001}, which the present analysis exploits. %

\section{Selection\label{sec:selection}}
The selection criteria are mainly inherited from the measurement of the difference between the decay-time-integrated \CP asymmetries in \Dkk and \Dpp decays~\cite{LHCb-PAPER-2019-006}, which uses the same sample of proton-proton collisions. Signal candidates are first required to pass the hardware trigger, which selects events containing at least one charged particle with high transverse momentum that leaves a track in the muon system. At the first stage of the software trigger, events are selected if they contain at least one track having large transverse momentum and being incompatible with originating from any PV, or if any two-track combination forming a secondary vertex passes a multivariate classifier. If a particle is identified as a muon, a lower \pt threshold is applied. At the second stage of the software trigger, the full event reconstruction is performed, and requirements on kinematic, topological, and particle-identification criteria are placed on the signal candidates. A \Dz candidate is formed by combining two well-reconstructed, oppositely charged tracks such that they are consistent with originating from a common vertex. The \Dz candidate must satisfy requirements on the vertex quality and has to be well separated from all PVs in the event. At the next step, the \Dz candidate is combined with a muon to form a \B candidate. Only candidates where the \Dz meson decays downstream along the beam axis with respect to the \B candidate are further considered. The \B candidate must have a visible mass, $m(\Dz\mu)$, and a corrected mass, $m_{\rm corr}(\B)$, consistent with a signal decay. %
The corrected mass is computed as $m_{\rm corr}(B)\equiv\sqrt{m^2(\Dz\mu) + p_\perp^2(\Dz\mu)} + p_\perp(\Dz\mu)$, where $p_\perp(\Dz\mu)$ is the momentum of the $\Dz\mu$ system transverse to the \B flight direction, to partially correct for the unreconstructed particles in the decay of the \B hadron.

In the off-line selection, trigger signals are associated with reconstructed particles. Particle-identification criteria and requirements on $m(\Dz\mu)$ and $m_{\rm corr}(\B)$ are tightened with respect to the on-line selection. The mass of the \Dz candidate is required to be in the ranges $[1825, 1925]\mevcc$, $[1820, 1939]\mevcc$ and $[1780, 1940]\mevcc$ for \mbox{\Dkk}, \mbox{\Dpp}, and \mbox{\Dkp} decays, respectively, to reduce the amount of background decays with misidentified final-state particles to a negligible level. The reconstructed decay time is computed from the distance, $L$,  between the measured \Dz and \B decay vertices and from the \Dz momentum, $p(\Dz)$, as $t = m_\Dz L/[p(\Dz) c]$, where $m_\Dz$ is the known \Dz mass~\cite{PDG2018}. All \Dz candidates with a reconstructed decay time that is either negative or exceeds 10 times the \Dz lifetime are discarded. Mass vetoes suppress background from misreconstructed \B decays to final states involving a charmonium resonance, such as $\B^- \to \psi^{(')}(\to \mu^+\mu^-)h^-$ with $h=\pi$ or $K$, where a muon is misidentified as a pion or kaon and is used in the \Dz final state. Tag muons reconstructed in regions of phase space with large instrumental asymmetries, due to muons of one charge either being bent out of the detector acceptance or deflected into the LHC beam pipe, are vetoed. The fraction of signal candidates removed by this requirement is $10\%$. In addition, for \Dkp decays, candidates with kaon $\pt<800\mevc$ are removed to reduce instrumental asymmetry between the detection of negatively and positively charged kaons. Since these requirements do not reduce the background to a sufficiently low level for \Dkk and \Dpp decays, a dedicated boosted decision tree (BDT) is trained to isolate the signal candidates from background made of accidental combinations of charged particles (``combinatorial background''). The variables used in the BDT to discriminate signal from combinatorial background are the fit quality of the \Dz and the \B decay vertices, the \Dz flight distance; the \Dz impact parameter with respect to the closest PV, the transverse momenta of the \Dz decay products, the significance of the distance between the \Dz and \B decay vertices, and the visible and corrected masses of the \B-hadron candidate. The BDT is trained using \Dkp decays as signal proxies and candidates from the \Dz mass sidebands of the signal decay modes as background. The optimal requirement on the BDT discriminant is chosen by maximizing the figure of merit $\mathcal{S}/\sqrt{\mathcal{S+B}}$ in a range corresponding to approximately 3 times the mass resolution around the \Dz mass, where $\mathcal{S}$ and $\mathcal{B}$ denote the signal and background yields, respectively. If an event contains more than one candidate after the full selection, one is chosen at random. The fraction of candidates removed by this requirement is $0.4\%$.

\begin{figure}[t]
\centering
  \begin{overpic}[width=0.5\linewidth]{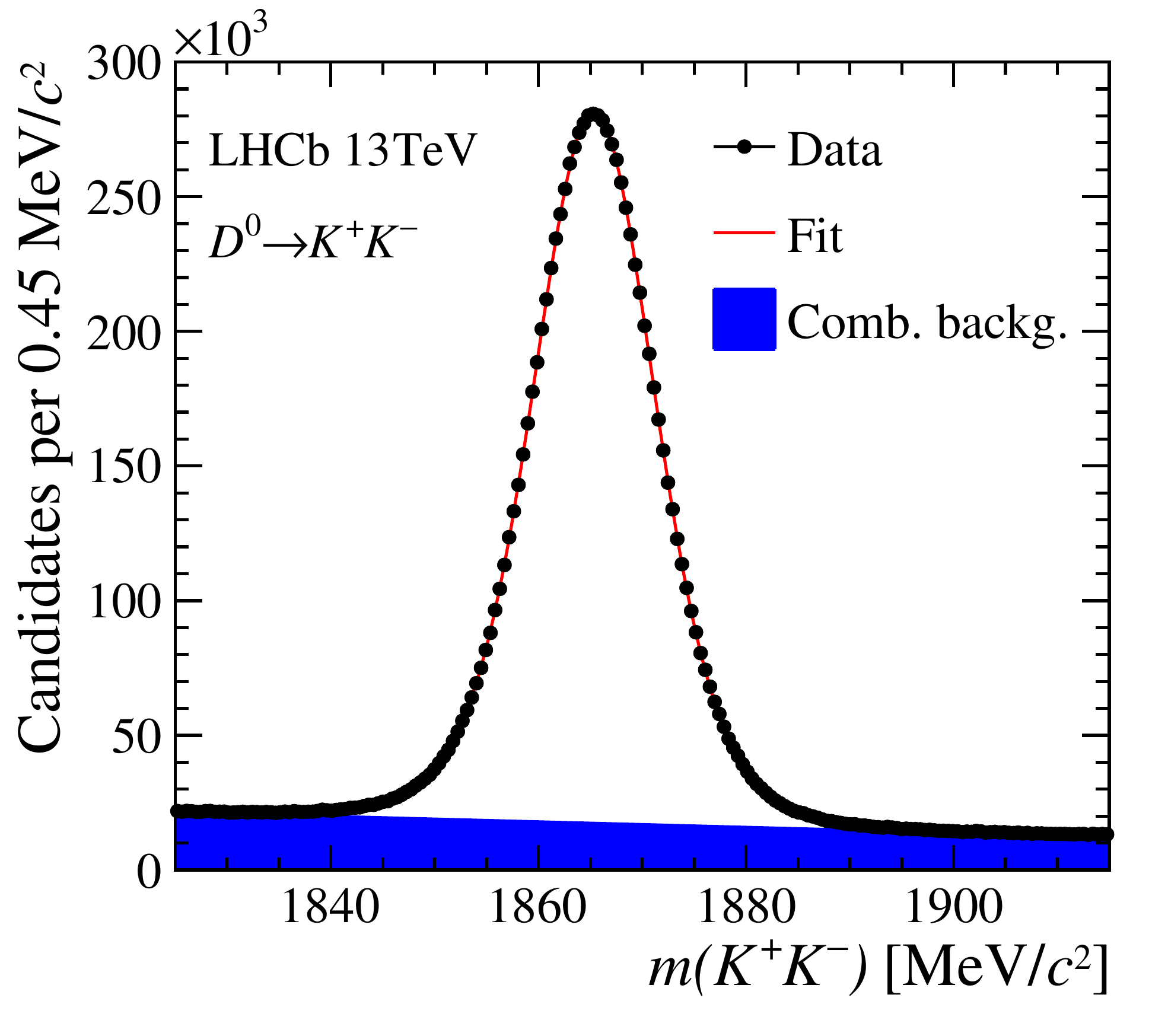}
      \put(21,55){$(a)$}
    \end{overpic}\hfil
      \begin{overpic}[width=0.5\linewidth]{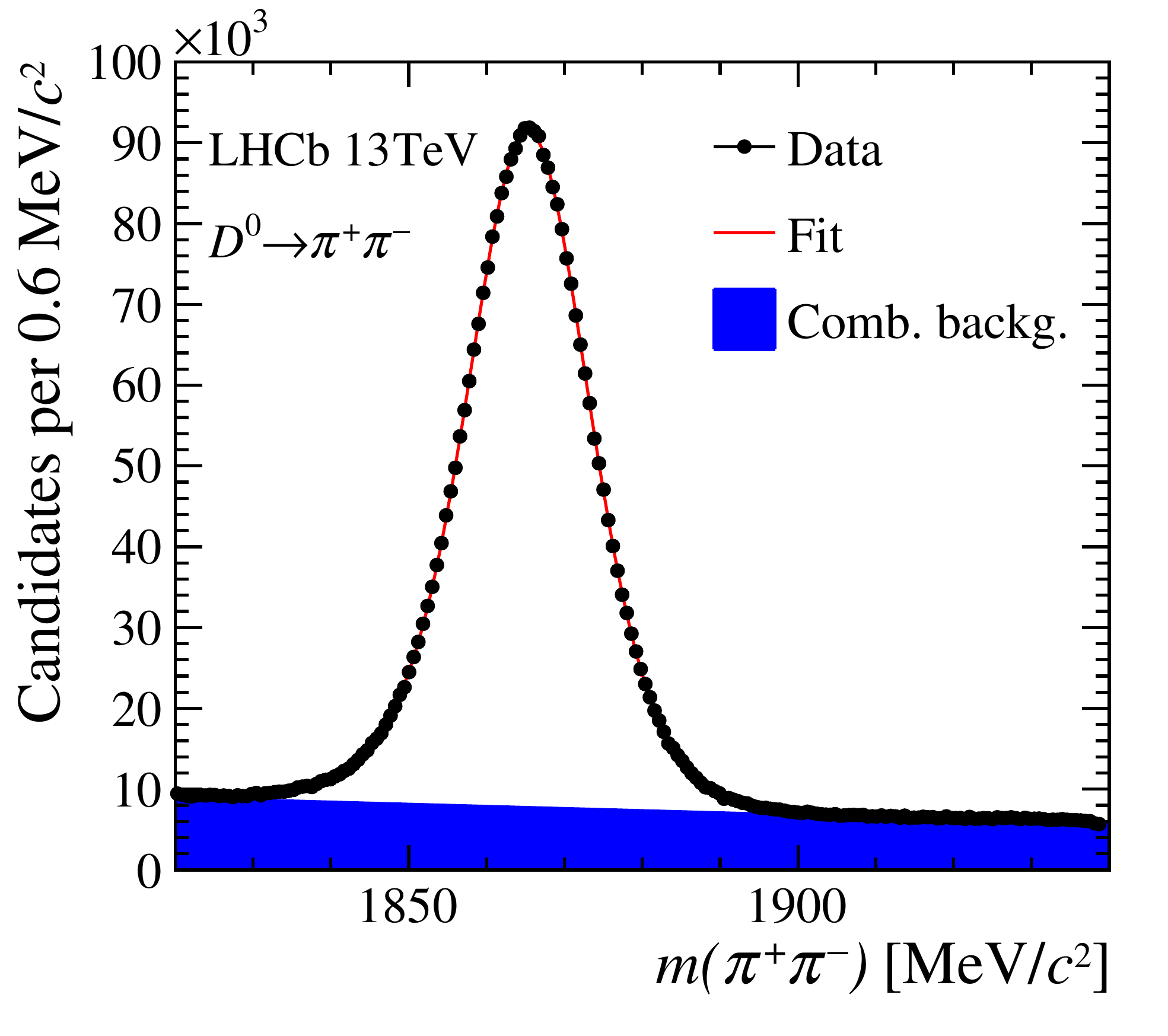}
      \put(21,55){$(b)$}
    \end{overpic}\\
      \begin{overpic}[width=0.5\linewidth]{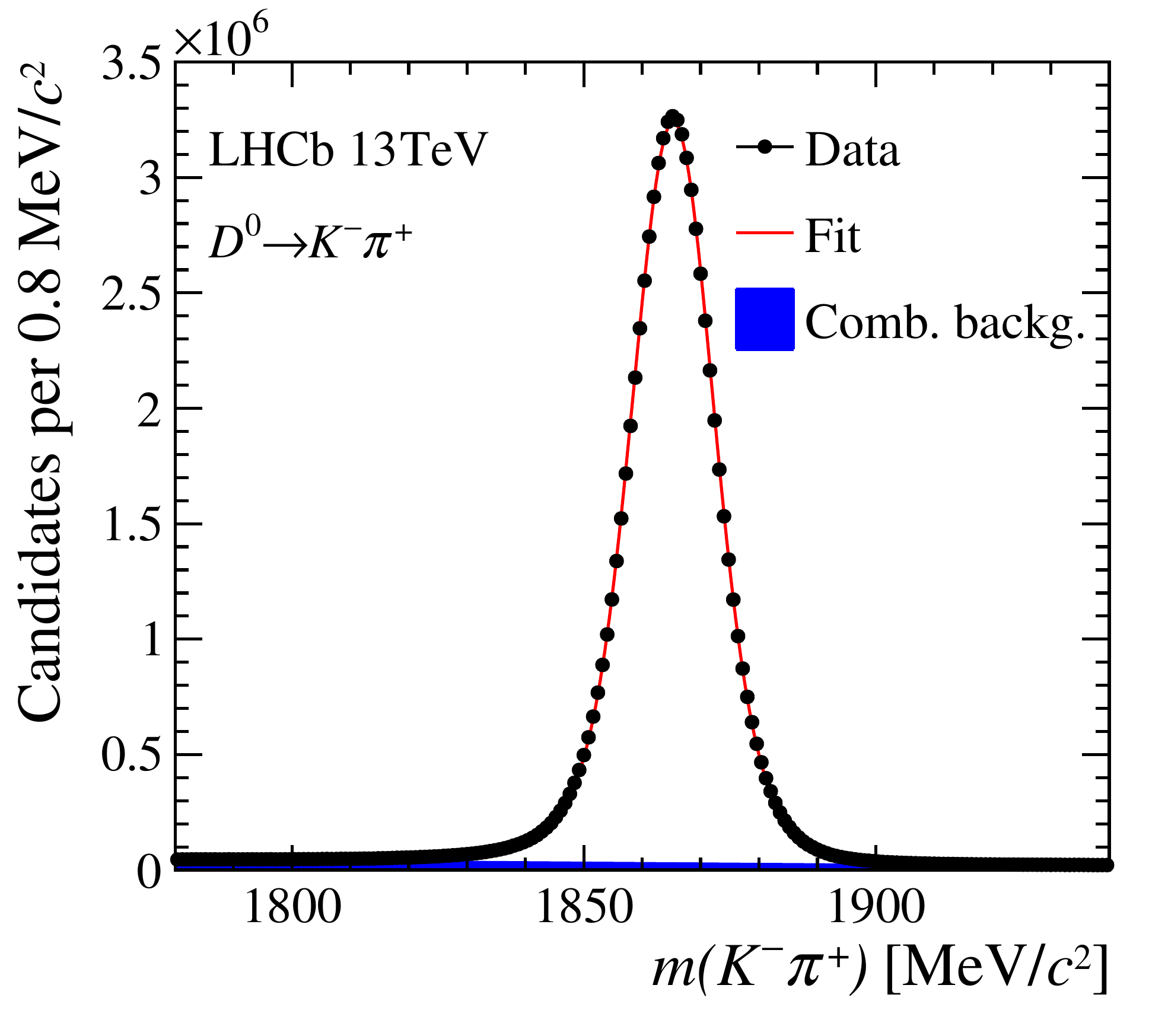}
      \put(21,55){$(c)$}
    \end{overpic}
   \caption{Mass distributions of (a) \Dkk, (b) \Dpp and (c) \Dkp candidates with fit projections overlaid.}
\label{fig:fit_mass} 
\end{figure} 

The mass distributions of the selected signal- and control-decay candidates are shown in Fig.~\ref{fig:fit_mass}. Details about the fit model are given in the next section. Approximately $9\times 10^6$, $3\times 10^6$, and $76\times 10^6$ signal \Dkk, \Dpp, and \Dkp decays, respectively, are  reconstructed over a smooth background dominated by accidental combinations of charged particles.

\section{Fit method\label{sec:fit}}
The samples of selected \Dkk, \Dpp and \Dkp candidates are split into 20 approximately equally populated subsets (``bins'') of decay time in the range $[0,10]\tau$. In each decay-time bin, the raw asymmetry \Araw is determined by a simultaneous binned $\chi^2$ fit to the \mD distributions of the \Dz and \Dzb candidates, split according to the muon tag. The total signal yields and asymmetries are treated as shared floating parameters of the fit. The fits include two components: signal and combinatorial background. The signal is described with a sum of a Gaussian and a Johnson's $S_U$ distribution~\cite{johnson}, with parameters determined from a fit to the decay-time-integrated mass spectra. To account for the observed dependence of the signal mass shape on decay time, the means and widths of the signal distributions are left free to float individually for each decay-time bin. The mass shape is assumed to be the same for \Dz and \Dzb candidates for charge-symmetric final states of the signal modes, and allowed to differ for \Dkp and $\Dzb\to\Kp\pim$ candidates. The combinatorial background is described by a linear function, with a slope that floats independently in each decay-time bin and is allowed to differ between \Dz and \Dzb candidates.

\begin{figure}[t]
\centering
\includegraphics[width=0.7\linewidth]{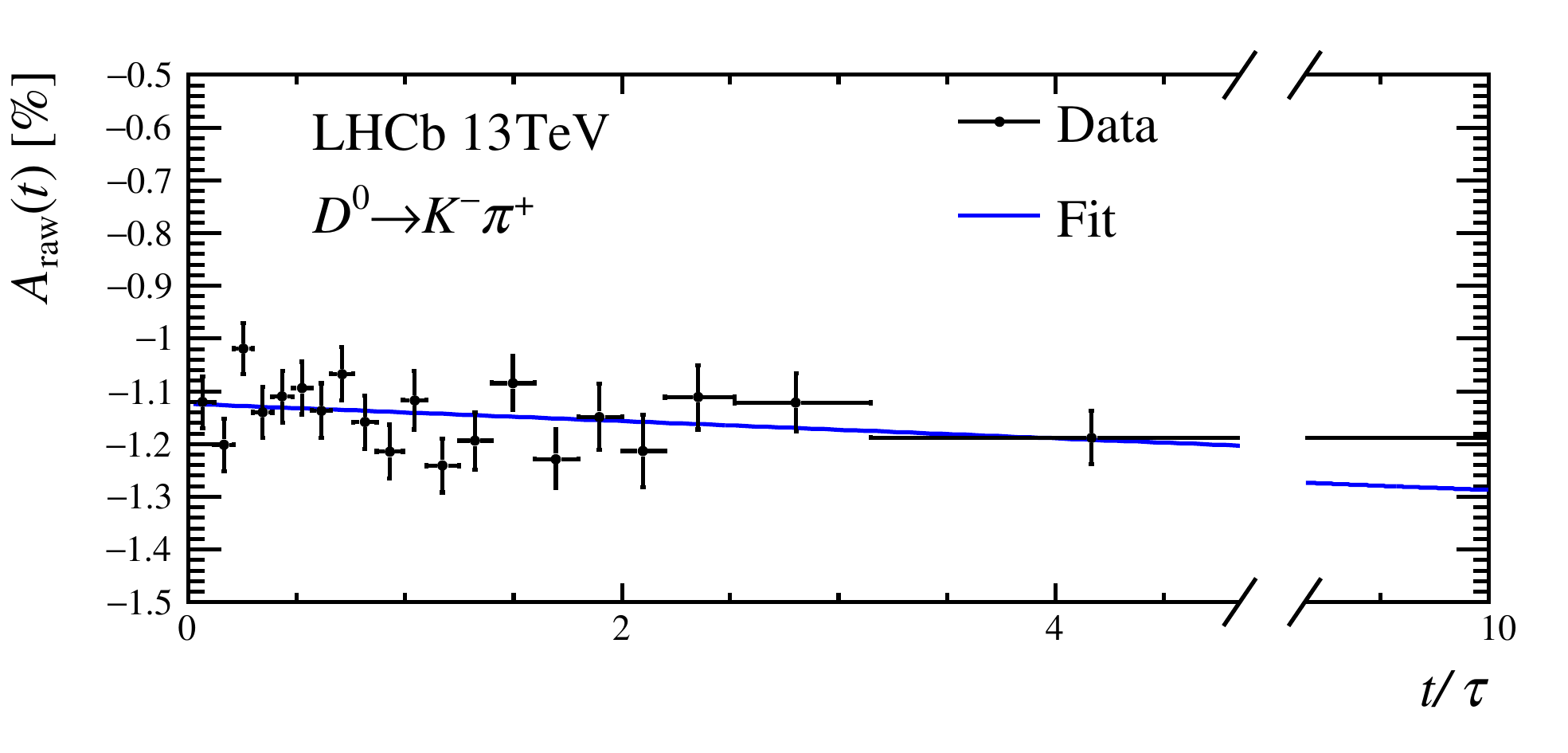}\\
\caption{Raw asymmetry as a function of decay time with fit projection overlaid for \Dkp signal candidates.\label{fig:fit_Agamma_kpi}}
\end{figure}

The raw asymmetry measured in decay-time bin $i$ is fit by minimizing the least squares with respect to the linear function $\Araw(0)-\agamma\langle t\rangle_i/\tau$. The decay-time-independent terms of Eqs.~\eqref{eq:approx-acp} and \eqref{eq:araw} are incorporated into a single parameter, $\Araw(0)$, that is determined by the fit together with \agamma. The average decay time in each bin $i$, $\langle t\rangle_i$, is computed using the decay-time distribution of background-subtracted \Dz candidates. Statistically consistent values are found for the control and signal modes.  The \Dz lifetime $\tau$ is set to its known value~\cite{PDG2018}. Using large samples of simulated experiments, it is verified that the analysis procedure leads to unbiased estimates of the fit parameters and of their uncertainties. Figure~\ref{fig:fit_Agamma_kpi} shows the projection of the decay-time-dependent fit to the \Dkp control sample. Here \agamma is measured to be $(\kpRes \pm \kpStat)\times\Units$, where the uncertainty is statistical only. The measured value is consistent with zero as expected, confirming the validity of the assumption of decay-time-independent nuisance asymmetries. In \Dkp decays, due to their charge-asymmetric final states, detection asymmetries are more pronounced compared to the signal modes, where these asymmetries are only caused by the muons used to tag the flavor of the \Dz mesons.    

\section{Systematic uncertainties\label{sec:systematics}}
The systematic uncertainty is dominated by the following contributions: the impact of decay-time acceptance and resolution, the effect of neglected background from combinations of real \Dz candidates with unrelated muons (which might lead to a wrong identification of the neutral \D-meson flavor), and the impact of the assumed parametrization of the signal and background mass shapes. These effects are studied using large samples of pseudoexperiments, where the above sources of systematic biases are simulated.

The average decay-time resolution is estimated to be $127\fs$ using simulated decays. In the generation of the pseudoexperiments, the resolution is increased by 10\% to account for differences between data and simulation. The decay-time acceptance is estimated from data by comparing the background-subtracted decay-time distributions of \Dkp candidates with an exponential function convoluted with the decay-time resolution. Different sets of pseudoexperiments, simulating the effect of decay-time acceptance and resolution, are generated with values of \agamma in the range $[-30,30]\times\Units$. Each pseudoexperiment is then fit with the default analysis approach, and the difference between the measured and the input values of \agamma is used to determine the systematic bias. As the bias is found to depend linearly on the true value of \agamma, the largest bias observed within the $68\%$ confidence-level interval of the current world average~\cite{HFLAV18} is taken as the systematic uncertainty. This amounts to $0.3\times\Units$ ($0.4\times\Units$) for \Dkk (\Dpp) decays.

The probability to wrongly associate unrelated muons with the \Dz candidates is estimated using the yields of ``wrong-sign'' $\Dz(\to K^-\pi^+)\mu^+$ and $\Dzb(\to K^+\pi^-)\mu^-$ candidates in data, which are corrected for the rate of doubly Cabibbo-suppressed decays and decays due to flavor oscillation using the measurements reported in Ref.~\cite{LHCb-PAPER-2017-046}. Mistag probabilities between 1\% at low decay times and 3\% at high decay times are observed. Also in this case the bias observed in pseudoexperiments depends linearly on the true value of \agamma. Following the same strategy as discussed above, a systematic uncertainty of $0.3\times\Units$ ($0.6\times\Units$) is assigned for \Dkk (\Dpp) decays.

To estimate any potential bias due to the specific choice of the mass model used in the fits that determine the raw asymmetries, samples of pseudoexperiments are generated using alternative signal and background models that describe the data equally well. The observed bias is independent of the input \agamma and results in an additional systematic uncertainty of $0.3\times\Units$ for both signal decay channels.

Uncertainties on $\langle t\rangle_i/\tau$ arising from relative misalignments of subdetectors and from the uncertainty on the input value of the \Dz lifetime~\cite{PDG2018} give negligible contributions. Furthermore, unexpected biases due to a possible decay-time dependence of the nuisance asymmetries and due to the selection procedure are investigated using the \Dkp control sample and/or by measuring \agamma in disjoint subsamples split by magnetic-field polarity, year of data taking, and kinematic variables of the \B hadron, \Dz meson and muon candidates. No unexpected variations are observed, and no additional systematic uncertainties are assigned.

\begin{table}[t]
\caption{\label{tab:systematics}Summary of the dominant contributions to the systematic uncertainty on $\agamma(\Kp\Km)$ and $\agamma(\pip\pim)$.}%
\centering
\begin{tabular}{lcc}
\toprule
Source of uncertainty                & $\agamma(\Kp\Km)$ [$10^{-4}$] & $\agamma(\pip\pim)$ [$10^{-4}$] \\
\midrule
Decay-time resolution and acceptance & $0.3$            & $0.4$\\ 
Mistag probability                   & $0.3$            & $0.6$\\
Mass-fit model			             & $0.3$            & $0.3$\\ 
\midrule
Total                                & \kkSyst          & \ppSyst\\ 
\bottomrule
\end{tabular}
\end{table}

A summary of the relevant systematic uncertainties is given in Table~\ref{tab:systematics}. The total systematic uncertainty is obtained by summing in quadrature the individual components and amounts to $0.5\times\Units$ and $0.8\times\Units$ for $\agamma(\Kp\Km)$ and $\agamma(\pip\pim)$, respectively.

\section{Results and conclusions\label{sec:results}}
\begin{figure}[t]
\centering
\begin{overpic}[width=0.7\linewidth]{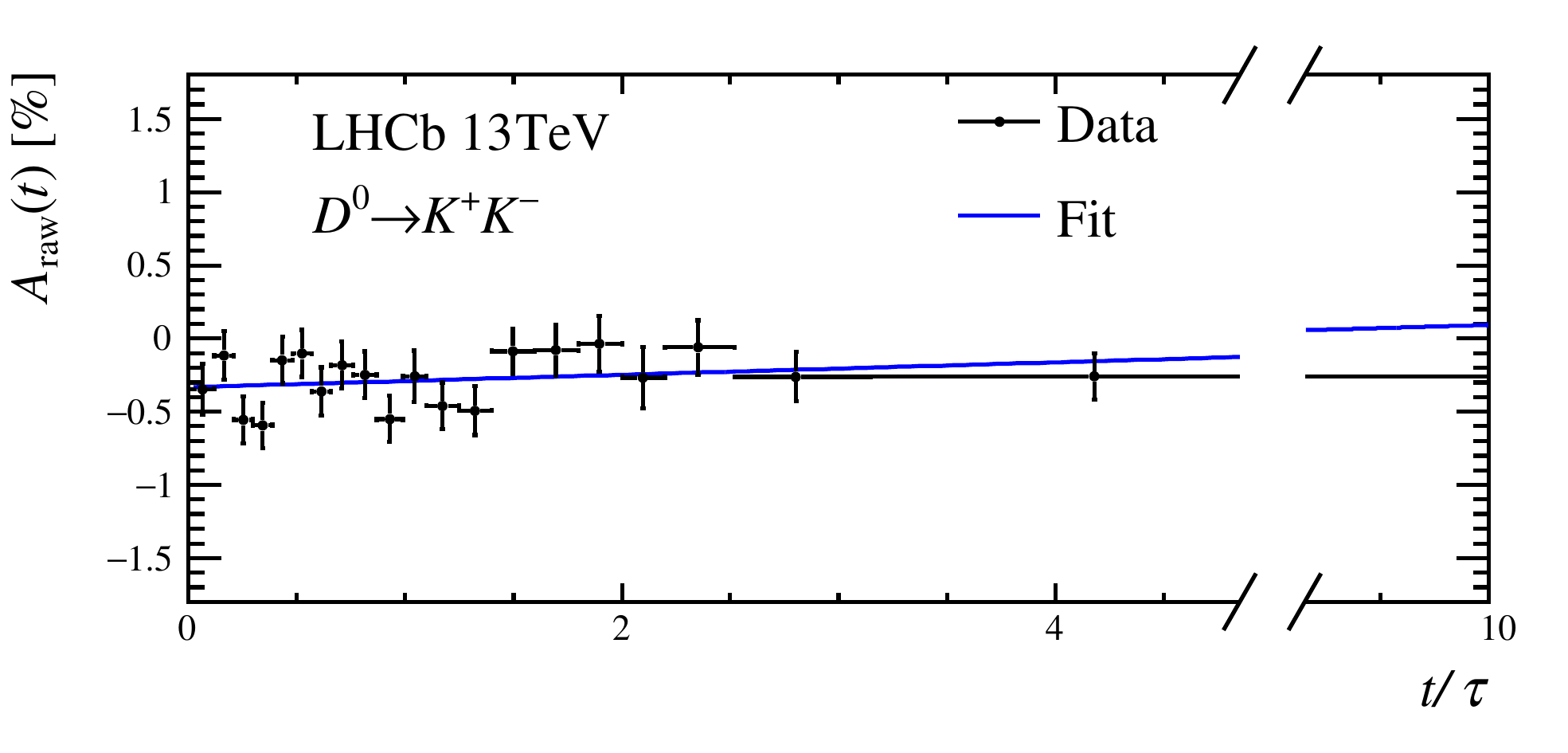}
\put(85,35){$(a)$} 
\end{overpic}\\
\begin{overpic}[width=0.7\linewidth]{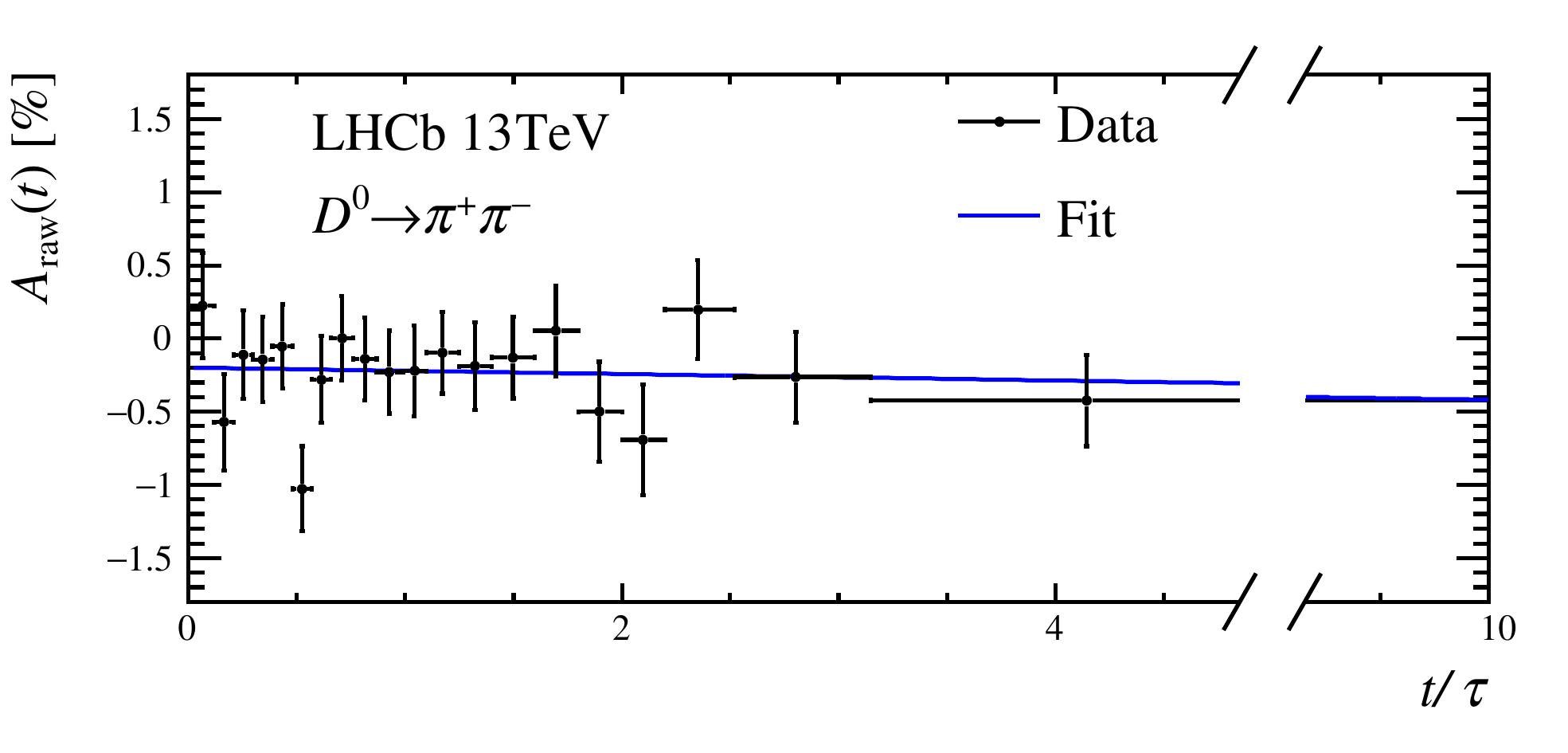}
\put(85,35){$(b)$}
\end{overpic}\\
\caption{Raw asymmetry as a function of decay time with fit projection overlaid for (a) \Dkk and (b) \Dpp signal candidates.}
\label{fig:fit_Agamma_hh}
\end{figure}

A search for decay-time-dependent \CP violation in \mbox{\Dkk} and \mbox{\Dpp} decays is performed using proton-proton collision data recorded with the \lhcb detector at a center-of-mass energy of 13\tev, and corresponding to an integrated luminosity of 5.4\invfb. The \Dz mesons are required to originate from semileptonic \bquark-hadron decays, such that the charge of the muon identifies the flavor of the neutral \D meson at the moment of its production. The parameter \agamma is determined from a fit to the asymmetry between \Dz and \Dzb yields as a function of decay time. The projections of the fits for both \Dkk and \Dpp samples are shown in Fig.~\ref{fig:fit_Agamma_hh}. The results are
\begin{align*}
\agamma(\Kp\Km) &= (\kkRes \pm \kkStat  \pm \kkSyst)\times\Units,\\
\agamma(\pip\pim) &= (\phantom{-}\ppRes \pm \ppStat  \pm \ppSyst)\times\Units,
\end{align*}
where the uncertainties are statistical and systematic, respectively.

The measured values are combined with previous \lhcb measurements based on data corresponding to 3\invfb collected at center-of-mass energies of 7 and 8\tev, and where the neutral \D mesons originate either from semileptonic \bquark-hadron decays~\cite{LHCb-PAPER-2014-069} or from promptly produced $\Dstarp(2010)$ mesons~\cite{LHCb-PAPER-2016-063}, with which they are consistent. The combination accounts for correlations in the systematic uncertainties and yields
\begin{align*}
\agamma(\Kp\Km) &= (\kkCombRes \pm \kkCombStat \pm \kkCombSyst)\times\Units,\\
\agamma(\pip\pim) &= (\phantom{-}\ppCombRes \pm \ppCombStat  \pm \ppCombSyst)\times\Units.
\end{align*}
Assuming \agamma to be universal, the above two results can be averaged to yield \mbox{$\agamma = (-2.9 \pm 2.0 \pm 0.6) \times\Units$}. The results do not show any indication of \CP violation in charm mixing or in the interference between mixing and decay.
 
%
%
\section*{Acknowledgements}
\noindent We express our gratitude to our colleagues in the CERN
accelerator departments for the excellent performance of the LHC. We
thank the technical and administrative staff at the LHCb
institutes.
We acknowledge support from CERN and from the national agencies:
CAPES, CNPq, FAPERJ and FINEP (Brazil); 
MOST and NSFC (China); 
CNRS/IN2P3 (France); 
BMBF, DFG and MPG (Germany); 
INFN (Italy); 
NWO (Netherlands); 
MNiSW and NCN (Poland); 
MEN/IFA (Romania); 
MSHE (Russia); 
MinECo (Spain); 
SNSF and SER (Switzerland); 
NASU (Ukraine); 
STFC (United Kingdom); 
DOE NP and NSF (USA).
We acknowledge the computing resources that are provided by CERN, IN2P3
(France), KIT and DESY (Germany), INFN (Italy), SURF (Netherlands),
PIC (Spain), GridPP (United Kingdom), RRCKI and Yandex
LLC (Russia), CSCS (Switzerland), IFIN-HH (Romania), CBPF (Brazil),
PL-GRID (Poland) and OSC (USA).
We are indebted to the communities behind the multiple open-source
software packages on which we depend.
Individual groups or members have received support from
AvH Foundation (Germany);
EPLANET, Marie Sk\l{}odowska-Curie Actions and ERC (European Union);
ANR, Labex P2IO and OCEVU, and R\'{e}gion Auvergne-Rh\^{o}ne-Alpes (France);
Key Research Program of Frontier Sciences of CAS, CAS PIFI, and the Thousand Talents Program (China);
RFBR, RSF and Yandex LLC (Russia);
GVA, XuntaGal and GENCAT (Spain);
the Royal Society
and the Leverhulme Trust (United Kingdom).

 
\addcontentsline{toc}{section}{References}
\bibliographystyle{LHCb}
\ifx\mcitethebibliography\mciteundefinedmacro
\PackageError{LHCb.bst}{mciteplus.sty has not been loaded}
{This bibstyle requires the use of the mciteplus package.}\fi
\providecommand{\href}[2]{#2}

\newpage
%
\centerline{\large\bf LHCb collaboration}
\begin{flushleft}
\small
R.~Aaij$^{28}$,
C.~Abell{\'a}n~Beteta$^{46}$,
T.~Ackernley$^{56}$,
B.~Adeva$^{43}$,
M.~Adinolfi$^{50}$,
H.~Afsharnia$^{6}$,
C.A.~Aidala$^{77}$,
S.~Aiola$^{22}$,
Z.~Ajaltouni$^{6}$,
S.~Akar$^{61}$
% ...

{\footnotesize \it
% ...
$ ^{6}$Universit{\'e} Clermont Auvergne, CNRS/IN2P3, LPC, Clermont-Ferrand, France\\
% ...
$ ^{22}$INFN Sezione di Milano, Milano, Italy\\
% ...
$ ^{28}$Nikhef National Institute for Subatomic Physics, Amsterdam, Netherlands\\
% ...
$ ^{43}$Instituto Galego de F{\'\i}sica de Altas Enerx{\'\i}as (IGFAE), Universidade de Santiago de Compostela, Santiago de Compostela, Spain\\
% ...
$ ^{46}$Physik-Institut, Universit{\"a}t Z{\"u}rich, Z{\"u}rich, Switzerland\\
% ...
$ ^{50}$H.H. Wills Physics Laboratory, University of Bristol, Bristol, United Kingdom\\
% ...
$ ^{56}$Oliver Lodge Laboratory, University of Liverpool, Liverpool, United Kingdom\\
% ...
$ ^{61}$University of Cincinnati, Cincinnati, OH, United States\\
% ...
$ ^{63}$Syracuse University, Syracuse, NY, United States\\
% ...
$ ^{77}$University of Michigan, Ann Arbor, United States, associated to $^{63}$\\
% ...
}
\end{flushleft} 
 
\end{document}